\def\year{2017}\relax
\documentclass[letterpaper]{article}
\usepackage{aaai17}
\usepackage{times}
\usepackage{helvet}
\usepackage{courier}
\usepackage{graphicx}
\usepackage{color}
\usepackage{url}
\usepackage{multirow}
\usepackage[flushleft]{threeparttable}

\nocopyright
\def\cluster#1{{\texttt{C#1}}} 

\begin{document}

\title{Online Human-Bot Interactions:  Detection, Estimation, and Characterization}

\author{
Onur Varol,\textsuperscript{1,*}
Emilio Ferrara,\textsuperscript{2} 
Clayton A. Davis,\textsuperscript{1}
Filippo Menczer,\textsuperscript{1}
Alessandro Flammini\textsuperscript{1}\\
\textsuperscript{1}Center for Complex Networks and Systems Research, Indiana University, Bloomington, US
\\
\textsuperscript{2}Information Sciences Institute, University of Southern California, Marina del Rey, CA, US}
\maketitle

\begin{abstract}

Increasing evidence suggests that a growing amount of social media content is generated by autonomous entities known as social bots. In this work we present a framework to detect such entities on Twitter. We leverage more than a thousand features extracted from public data and meta-data about users: friends, tweet content and sentiment, network patterns, and activity time series. We benchmark the classification framework by using a publicly available dataset of Twitter bots. This training data is enriched by a manually annotated collection of active Twitter users that include both humans and bots of varying sophistication. Our models yield high accuracy and agreement with each other and can detect bots of different nature. Our estimates suggest that between 9\% and 15\% of active Twitter accounts are bots. Characterizing ties among accounts, we observe that simple bots tend to interact with bots that exhibit more human-like behaviors. Analysis of content flows reveals retweet and mention strategies adopted by bots to interact with different target groups. Using clustering analysis, we characterize several subclasses of accounts, including spammers, self promoters, and accounts that post content from connected applications.
\end{abstract}

\section{Introduction}

Social media are powerful tools connecting millions of people across the globe. These connections form the substrate that supports information dissemination, which ultimately affects the ideas, news, and opinions to which we are exposed. 
There exist entities with both strong motivation and technical means to abuse online social networks --- from individuals aiming to artificially boost their popularity, to organizations with an agenda to influence public opinion. 
It is not difficult to automatically target particular user groups and promote specific content or views~\cite{ferrara2014rise,bessi2016social}. 
Reliance on social media may therefore make us vulnerable to manipulation.

\textit{Social bots} are accounts controlled by software, algorithmically generating  content and establishing interactions. Many social bots perform useful functions, such as dissemination of news and publications~\cite{lokot2016news,haustein2016tweets} and coordination of volunteer activities~\cite{savage2016botivist}.
However, there is a growing record of malicious applications of social bots.  Some emulate human behavior to manufacture fake grassroots political support~\cite{ratkiewicz2011detecting}, promote terrorist propaganda and recruitment~\cite{berger2015isis,abokhodair2015dissecting,ferrara2016predicting}, manipulate the stock market~\cite{ferrara2014rise}, and disseminate rumors and conspiracy theories~\cite{bessi2015science}.  

A growing body of research is addressing social bot activity, its implications on the social network, and the detection of these accounts~\cite{lee2011seven,boshmaf2011socialbot,beutel2013copycatch,yang2014uncovering,ferrara2014rise,Chavoshi2016}. The magnitude of the problem was underscored by a Twitter bot detection challenge recently organized by DARPA to study information dissemination mediated by automated accounts and to detect malicious activities carried out by these bots~\cite{subrahmanian2016darpa}. 

\subsection*{Contributions and Outline}

Here we demonstrate that accounts controlled by software exhibit  behaviors that reflects their intents and \textit{modus operandi}~\cite{bakshy2011everyone,hin2016}, and that such behaviors can be detected by supervised machine learning techniques. 
This paper makes the following contributions:

\begin{itemize}
	\item We propose a framework to extract a large collection of features from data and meta-data about social media users, including friends, tweet content and sentiment, network patterns, and activity time series. We use these features to train highly-accurate models to identify bots. For a generic user, we produce a $[0,1]$ score representing the likelihood that the user is a bot. 
	
	\item The performance of our detection system is evaluated against both an existing public dataset and an additional sample of manually-annotated Twitter accounts collected with a different strategy. We enrich the previously-trained models using the new annotations, and investigate the effects of different datasets and classification models.
	
	\item We classify a sample of millions of English-speaking active users. We use different models to infer thresholds in the bot score that best discriminate between humans and bots. We estimate that the percentage of Twitter accounts exhibiting social bot behaviors is between 9\% and 15\%.
    
    \item We characterize friendship ties and information flow between users that show behaviors of  different nature: human and bot-like. 
    Humans tend to interact with more human-like accounts than bot-like ones, on average. Reciprocity of  friendship ties is higher for humans. Some bots target users more or less randomly, others can choose targets based on their intentions.
	
	\item Clustering analysis reveals certain specific behavioral groups of accounts. Manual investigation of samples extracted from each cluster points to three distinct bot groups: spammers, self promoters, and accounts that post content from connected applications.

\end{itemize}

\section{Bot Detection Framework}

In the next section, we introduce a Twitter bot detection framework (\url{truthy.indiana.edu/botornot}) that is freely available online. This system leverages more than one thousand features to evaluate the extent to which a Twitter account exhibits similarity to the known characteristics of social bots~\cite{davis2016botornot}.

\subsection{Feature Extraction}

Data collected using the Twitter API are distilled in 1,150 features in six different classes. The classes and types of features are reported in Table~\ref{tab:features} and discussed next. 

\begin{table*}[t!]
\centering
\caption{List of 1150 features extracted by our framework.} 
\small
\begin{threeparttable}
\begin{tabular}{@{}l@{}l@{}|l@{}l@{}}
\hline\multirow{20}{*}{\rotatebox{90}{\textbf{User meta-data}}}
& Screen name length & \multirow{18}{*}{\rotatebox{90}{\textbf{Sentiment}}} 
& (***) Happiness scores of aggregated tweets \\
& Number of digits in screen name & & (***) Valence scores of aggregated tweets \\
& User name length & & (***) Arousal scores of aggregated tweets\\
& Time offset (sec.) & & (***) Dominance scores of single tweets \\
& Default profile (binary) & & (*) Happiness score of single tweets\\
& Default picture (binary) & & (*) Valence score of single tweets\\
& Account age (days) & & (*) Arousal score of single tweets\\
& Number of unique profile descriptions & & (*) Dominance score of single tweets\\
& (*) Profile description lengths & & (*) Polarization score of single tweets\\
& (*) Number of friends distribution & & (*) Entropy of polarization scores of single tweets\\
& (*) Number of followers distribution & & (*) Positive emoticons entropy of single tweets \\
& (*) Number of favorites distribution & & (*) Negative emoticons entropy of single tweets \\
& Number of friends (signal-noise ratio and rel. change) & & (*) Emoticons entropy of single tweets \\
& Number of followers (signal-noise ratio and rel. change) & & (*) Positive and negative score ratio of single tweets \\
& Number of favorites (signal-noise ratio and rel. change) & & (*) Number of positive emoticons in single tweets \\
& Number of tweets (per hour and total) & & (*) Number of negative emoticons in single tweets \\
& Number of retweets (per hour and total) & & (*) Total number of emoticons in single tweets \\
& Number of mentions (per hour and total) & & Ratio of tweets that contain emoticons\\
& Number of replies (per hour and total) & & \\
& Number of retweeted (per hour and total) & & \\

\hline\multirow{10}{*}{\rotatebox{90}{\textbf{Friends ($\dag$)}}}
& Number of distinct languages & 
\multirow{10}{*}{\rotatebox{90}{\textbf{Network ($\ddagger$)}}}
& Number of nodes\\
& Entropy of language use & & Number of edges (also for reciprocal) \\ 
& (*) Account age distribution & & (*) Strength distribution  \\
& (*) Time offset distribution & & (*) In-strength distribution \\
& (*) Number of friends distribution & & (*) Out-strength distribution \\
& (*) Number of followers distribution & & Network density (also for reciprocal) \\
& (*) Number of tweets distribution & & (*) Clustering coeff. (also for reciprocal) \\
& (*) Description length distribution & & \\
& Fraction of users with default profile and default picture & & \\

\hline\multirow{4}{*}{\rotatebox{90}{\textbf{Content}}} 
& (*,**) Frequency of POS tags in a tweet &
\multirow{4}{*}{\rotatebox{90}{\textbf{Timing}}} 
& (*) Time between two consecutive tweets \\
& (*,**) Proportion of POS tags in a tweet & & (*) Time between two consecutive retweets\\
& (*) Number of words in a tweet & & (*) Time between two consecutive mentions \\
& (*) Entropy of words in a tweet & & \\

\hline
\end{tabular} 
\begin{tablenotes}
\item[$\dag$] We consider four types of connected users: retweeting, mentioning, retweeted, and mentioned.
\item[$\ddagger$] We consider three types of network: retweet, mention, and hashtag co-occurrence networks. 
\item[*] Distribution types. For each distribution, the following eight statistics are computed and used as individual features: min, max, median, mean, std. deviation, skewness, kurtosis, and entropy. 
\item[**] Part-Of-Speech (POS) tag. There are nine POS tags: verbs, nuns, adjectives, modal auxiliaries, pre-determiners, interjections, adverbs, wh-, and pronouns.
\item[***] For each feature, we compute mean and std. deviation of the weighted average across words in the lexicon. 
\bigskip
\end{tablenotes}
\end{threeparttable}
\label{tab:features}
\end{table*}

\subsubsection{User-based features.}

Features extracted from user meta-data have been used to classify users and patterns before~\cite{mislove2011understanding,ferrara2014rise}.
We extract user-based features from meta-data available through the Twitter API. Such features include the number of friends and followers, the number of tweets produced by the users, profile description and settings.

\subsubsection{Friends features.}

Twitter actively fosters inter-connectivity. Users are linked by follower-friend (followee) relations. Content travels from person to person via retweets. Also, tweets can be addressed to specific users via mentions. We consider four types of links: retweeting, mentioning, being retweeted, and being mentioned. For each group separately, we extract features about language use, local time, popularity, etc.
Note that, due to Twitter's API limits, we do not use follower/followee information beyond these aggregate statistics.

\subsubsection{Network features.}

The network structure carries crucial information for the characterization of different types of communication. In fact, the usage of network features significantly helps in tasks like political astroturf detection \cite{ratkiewicz2011detecting}. Our system reconstructs three types of networks: retweet, mention, and hashtag co-occurrence networks. Retweet and mention networks have users as nodes, with a directed link between a pair of users that follows the direction of information spreading: toward the user retweeting or being mentioned.
Hashtag co-occurrence networks have undirected links between hashtag nodes when two hashtags occur together in a tweet. 
All networks are weighted according to the frequency of interactions or co-occurrences. 
For each network, we compute a set of features, including in- and out-strength (weighted degree) distributions, density, and clustering. Note that out-degree and out-strength are measures of popularity.

\subsubsection{Temporal features.}

Prior research suggests that the temporal signature of content production and consumption may reveal important information about online campaigns and their evolution~\cite{Ghosh11snakdd,ferrara2016campaign,Chavoshi2016}. To extract this signal we measure several temporal features related to user activity, including average rates of tweet production over various time periods and distributions of time intervals between events.

\subsubsection{Content and language features.}

Many recent papers have demonstrated the importance of content and language features in revealing the nature of social media conversations~\cite{danescu2013no,mcauley2013amateurs,mocanu2013twitter,botta2015quantifying,letchford2015advantage,hin2016}. For example, deceiving messages generally exhibit informal language and short sentences \cite{briscoe2014cues}. Our system does not employ features capturing the quality of tweets, but collects statistics about length and entropy of tweet text. Additionally, we extract language features by applying the \emph{Part-of-Speech} (POS) tagging technique, which identifies different types of natural language components, or \emph{POS tags}.
Tweets are therefore analyzed to study how POS tags are distributed. 

\subsubsection{Sentiment features.}

Sentiment analysis is a powerful tool to describe the emotions conveyed by a piece of text, and more broadly the attitude or mood of an entire conversation.
Sentiment extracted from social media conversations has been used to forecast offline events including
financial market fluctuations~\cite{bollen2011twitter}, and is known to affect information spreading \cite{mitchell2013geography,ferrara2015quantifying}. 
Our framework leverages several sentiment extraction techniques to generate various sentiment features, including \emph{arousal}, \emph{valence} and \emph{dominance} scores~\cite{warriner2013norms},  \emph{happiness} score~\cite{kloumann2012positivity}, \emph{polarization} and \emph{strength}~\cite{wilson2005recognizing}, and \emph{emoticon} score~\cite{agarwal2011sentiment}.

\subsection{Model Evaluation}
\label{sec:model-evaluation}

To train our system we initially used a publicly available dataset consisting of 15K manually verified Twitter bots  identified via a \textit{honeypot} approach~\cite{lee2011seven} and 16K verified human accounts. We collected the most recent tweets produced by those accounts using the Twitter Search API. We limited our collection to 200 public tweets from a user timeline and up to 100 of the most recent public tweets mentioning that user. This procedure yielded a dataset of 2.6 million tweets produced by manually verified  bots and 3 million tweets produced by human users. 

We benchmarked our system using several off-the-shelf algorithms provided in the \textit{scikit-learn} library~\cite{scikit-learn}. 
In a generic evaluation experiment, the classifier under examination is provided with  numerical vectors, each describing the features of an account. The classifier returns a numerical score in the unit interval. A higher score indicates a stronger belief that the account is a bot. 
A model's accuracy is evaluated by measuring the Area Under the receiver operating characteristic Curve (AUC) with 5-fold cross validation, and computing the average AUC score across the folds using Random Forests, AdaBoost, Logistic Regression and Decision Tree classifiers.
The best classification performance of 0.95 AUC was obtained by the \textit{Random Forest} algorithm. In the rest of the paper we use the Random Forest model trained using 100 estimators and the Gini coefficient to measure the quality of splits.

\section{Large-Scale Evaluation}

We realistically expect that the nature and sophistication of bots evolves over time and changes in specific conversational domains. It is therefore important to determine how reliable and consistent are the predictions produced by a system trained on a dataset but tested on  different data (in the wild). Also, the continuously-evolving nature of bots dictates the need to constantly update the models based on newly available training data.

To obtain an updated evaluation of the accuracy of our model, we constructed an additional, manually-annotated collection of Twitter user accounts.
We hypothesize that this recent collection includes some bots that are more sophisticated than the ones obtained years earlier with the honeypot method.
We leveraged these manual annotations to evaluate the model trained using the honeypot dataset and then to update the classifier's training data, producing a \textit{merged} dataset to train a new model that ensures better generalization to more sophisticated accounts. User IDs and annotation labels in our extended dataset 
are publicly available (\url{truthy.indiana.edu/botornot/data}).

\subsection{Data Collection}

Our data collection focused on users producing content in English, as inferred from profile meta-data. We identified a large, representative sample of users by monitoring a Twitter stream, accounting for approximately 10\% of public tweets, for 3 months starting in October 2015. This approach avoids known biases of other methods such as snowball and breadth-first sampling, which rely on the selection of an initial group of users~\cite{gjoka2010walking,morstatter2013sample}. 
We focus on English speaking users as they represent the largest group on Twitter~\cite{mocanu2013twitter}.

To restrict our sample to recently active users, we introduce the further criteria that they must have produced at least 200 tweets in total and 90 tweets during the three-month observation window (one per day on average). Our final sample includes approximately 14 million user accounts that meet both criteria. 
For each of these  accounts, we collected their tweets through the Twitter Search API.
We restricted the collection to the most recent 200 tweets and 100 mentions of each user, as described earlier. Owing to Twitter API limits, this greatly improved our data collection speed. 
This choice also reduces the response time of our service and API.
However the limitation adds noise to the features, due to the scarcity of data available to compute them. 

\subsection{Manual Annotations}
\label{sec:annotations}

We computed classification scores for each of the active accounts using our initial classifier trained on the honeypot dataset. 
We then grouped accounts by their bot scores, allowing us to evaluate our system across the spectrum of human and bot accounts without being biased by the distribution of bot scores.
We randomly sampled 300 accounts from each bot-score decile. The resulting balanced set of 3000 accounts were manually annotated by inspecting their public Twitter profiles. 
Some accounts have obvious flags, such as using a stock profile image or retweeting every message of another account within seconds. In general, however, there is no simple set of rules to assess whether an account is human or bot.  
With the help of four volunteers, we analyzed profile appearance, content produced and retweeted, and interactions with other users in terms of retweets and mentions. 
Annotators were not given a precise set of instructions to perform the classification task, but rather shown a consistent number of both positive and negative examples. The final decisions reflect each annotator's opinion and are restricted to: \textit{human, bot}, or \textit{undecided}. Accounts labeled as undecided were eliminated from further analysis. 

We annotated all 3000 accounts. We will refer to this set of accounts as the \emph{manually annotated} data set. Each annotator was assigned a random sample of accounts from each decile. We enforced a minimum 10\% overlap between annotations to assess the reliability of each annotator. This yielded an average pairwise agreement of 75\% and moderate inter-annotator agreement (Cohen's $\kappa=0.41$). We also computed the agreement between annotators and classifier outcomes, assuming that a classification score above 0.5 is interpreted as a bot. This resulted in an average pairwise agreement of 79\% and a moderately high Cohen's $\kappa=0.5$. 
These results suggest high confidence in the annotation process, as well as in the agreement between annotations and model predictions.

\subsection{Evaluating Models Using Annotated Data}

To evaluate our classification system trained on the honeypot dataset, we examined the classification accuracy separately for each bot-score decile of the \emph{manually annonated} dataset. 
We achieved classification accuracy greater than 90\% for the accounts in the $(0.0, 0.4)$ range, which includes mostly human accounts. We also observe accuracy above 70\% for scores in the $(0.8, 1.0)$ range (mostly bots). Accuracy for accounts in the grey-area range $(0.4, 0.8)$ fluctuates between 60\% and 80\%. Intuitively, this range contains the most challenging accounts to label, as reflected also in the low inter-annotators overlap in this region. When the accuracy of each bin is weighted by the population density in the large dataset from which the \emph{manually annonated} has been extracted, we obtain 86\% overall classification accuracy. 

We also compare annotator agreement scores for the accounts in each bot-score decile. We observe that agreement scores are higher for accounts in the $(0.0, 0.4)$ range and lower for accounts in the $(0.8, 1.0)$ range, indicating that it is more difficult for human annotators to identify bot-like as opposed to human-like behavior. 

We observe a similar pattern for the amount of time required on average to annotate human and bot accounts. Annotators employed on average 33 seconds to label human accounts and 37 seconds for bot accounts. 

\begin{figure}[!t]
	\centering
	\includegraphics[width=\columnwidth]{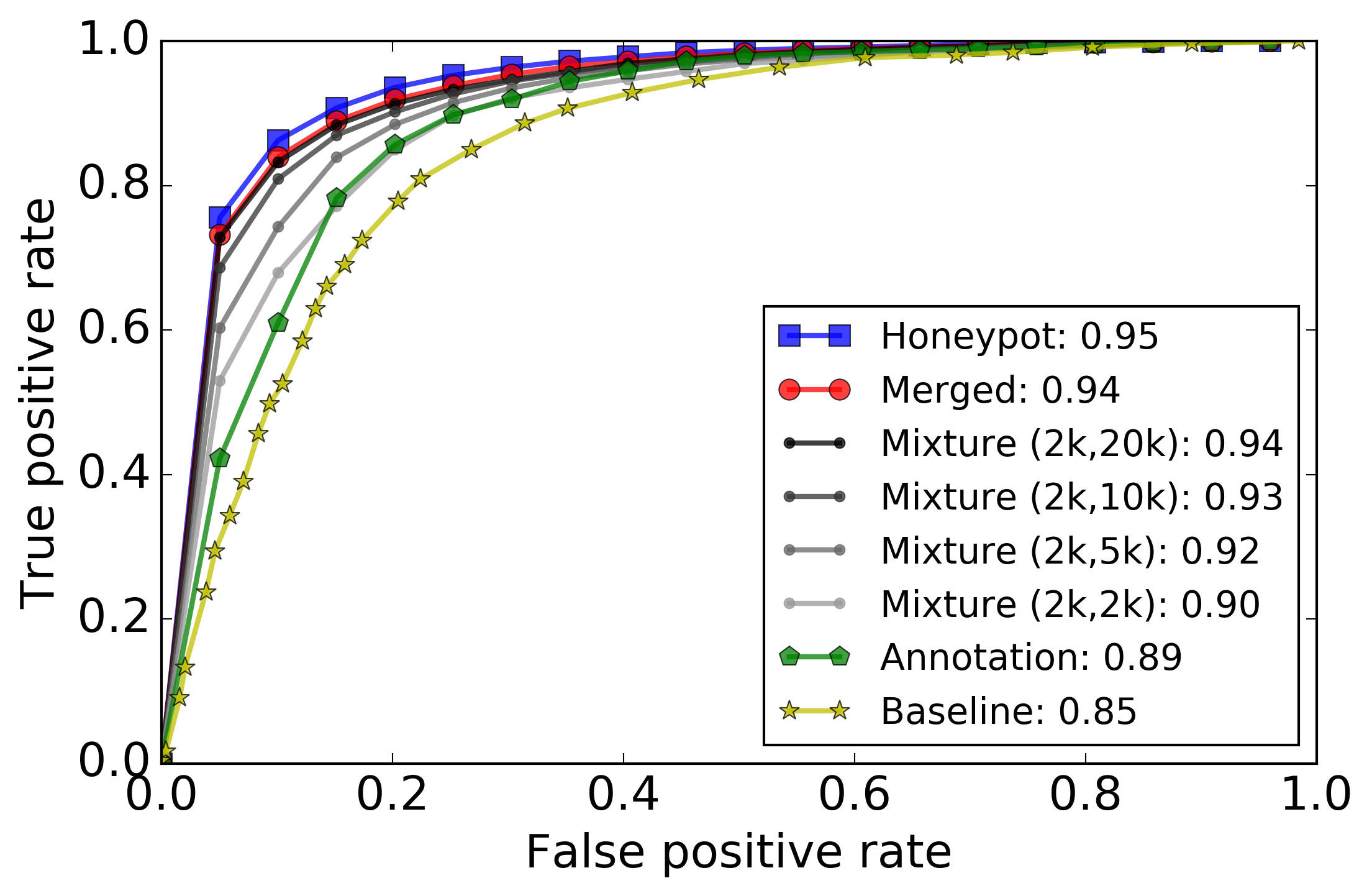}
	\caption{ROC curves of models trained and tested on different datasets. Accuracy is measured by AUC.}
	\label{fig:bot_detection_comparison}
\end{figure}

Fig.~\ref{fig:bot_detection_comparison} shows the results of experiments designed to investigate our ability to detect manually annotated bots. The baseline ROC curve is obtained by testing the honeypot model  on the manually annotated dataset. Unsurprisingly, the baseline accuracy (0.85 AUC) is lower than that obtained cross-validating on the honeypot data (0.95 AUC), because the model is not trained on the newer bots. 
 
\subsection{Dataset Effect on Model Accuracy}
\label{sec:dataset_effect}

We can update our models by combining the \emph{manually-annotated} and honeypot datasets.
We created multiple balanced datasets and performed 5-fold cross-validation to evaluate the accuracy of the corresponding models:

\begin{itemize}

\item \textbf{Annotation}: We trained and tested a model by only using annotated accounts and labels assigned by the majority of annotators. This yields 0.89 AUC, a reasonable accuracy considering that the dataset contains recent and possibly sophisticated bots.

\item \textbf{Merged}: We merged the honeypot and annotation datasets for training and testing. The resulting classifier achieves 0.94 AUC, only slightly worse than the honeypot (training and test) model although the \emph{merged} dataset contains a variety of more recent bots. 

\item \textbf{Mixture}: Using mixtures with different ratios of accounts from the \emph{manually annotated} and honeypot datasets, we obtain an accuracy ranging between 0.90 and 0.94 AUC.

\end{itemize}

\begin{figure}[!t]
	\centering
	\includegraphics[width=0.95\columnwidth]{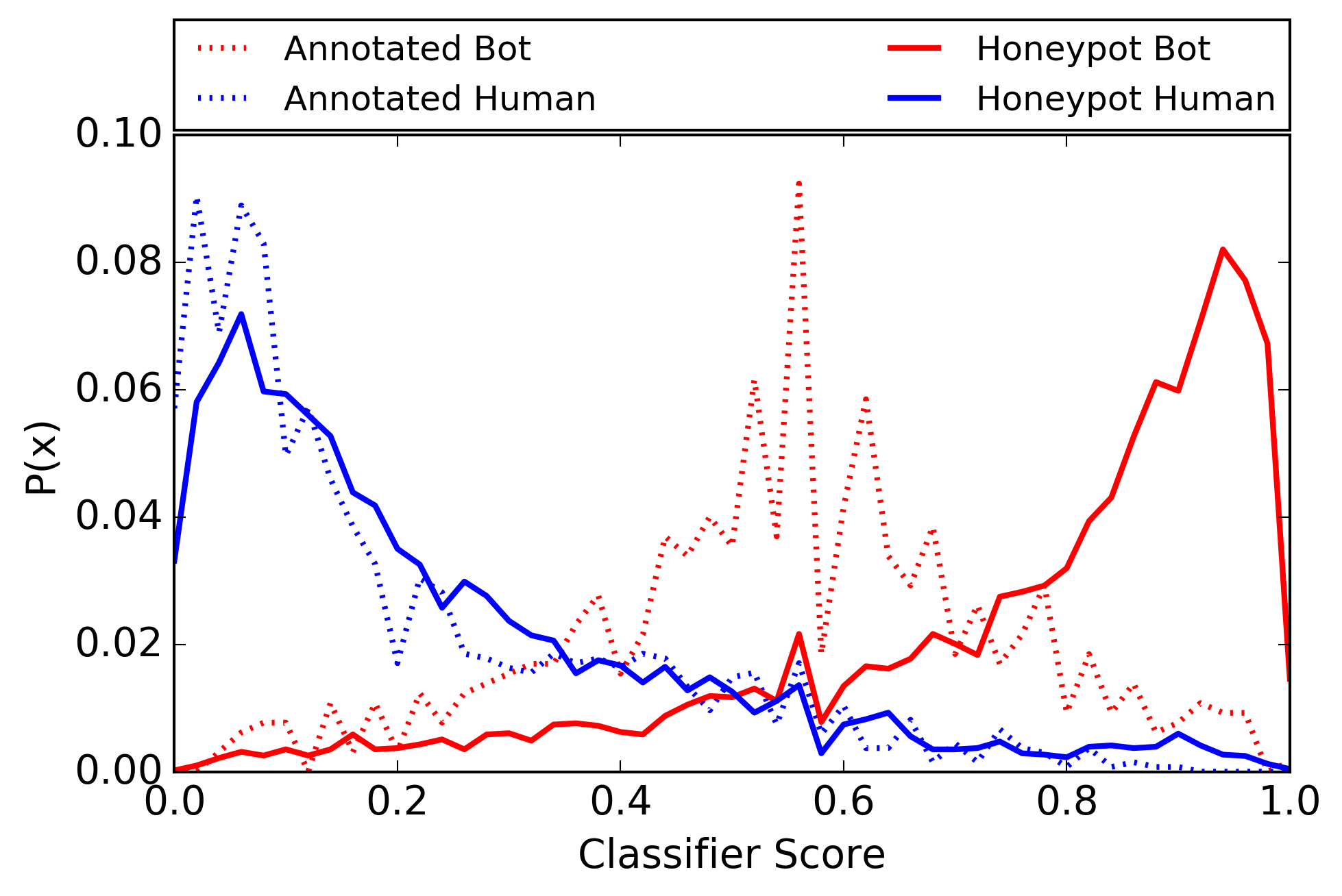}
	\caption{Distribution of classifier score for human and bot accounts in the two datasets.}
	\label{fig:dataset_effect_distributions}
\end{figure}

In Fig~\ref{fig:dataset_effect_distributions}, we plot the distributions of classification scores for human and bot accounts according to each dataset. The mixture model trained on 2K annotated and 10K honeypot accounts is used to compute the scores. Human accounts in both datasets have similar distributions, peaked around 0.1. The difference between bots in the two datasets is more prominent. The distribution of simple, honeypot bots peaks around 0.9. The newer bots from the \emph{manually annotated} dataset have typically smaller scores, with a distribution peaked around 0.6. They are more sophisticated, and exhibit characteristics more similar to human behavior. 
This raises the issue of how to properly set a threshold on the score when a strictly binary classification between human and bots is needed. To infer a suitable threshold, we compute classification accuracies for varying thresholds considering all accounts scoring below each threshold as human, and then select the threshold that maximizes accuracy.

\begin{figure}[!t]
	\centering
	\includegraphics[width=0.49\columnwidth]{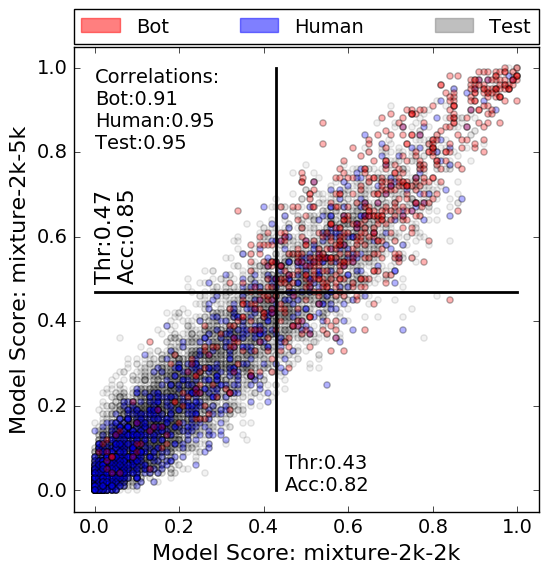}
	\includegraphics[width=0.49\columnwidth]{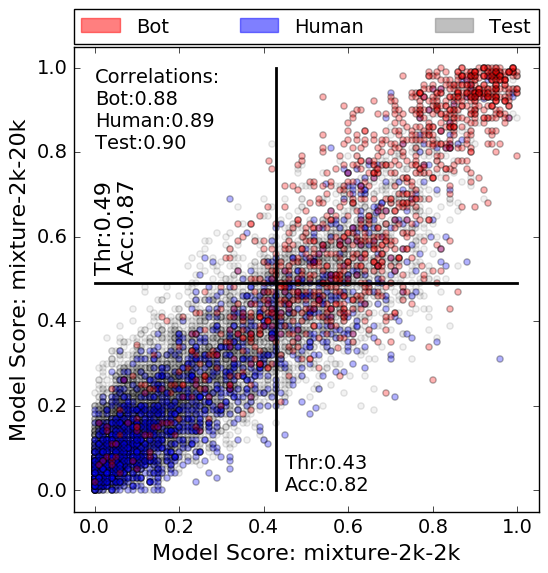}
	\caption{Comparison of scores for different models. Each account is represented as a point in the scatter plot with a color determined by its category. Test points are randomly sampled from our large-scale collection. Pearson correlations between scores are also reported, along with estimated thresholds and corresponding accuracies.}
	\label{fig:model_comparison_scatter}
\end{figure}

We compared scores for accounts in the \emph{manually annotated} dataset by pairs of models (\emph{i.e.} trained with different mixtures) for labeled human, bot, and a random subset of accounts (Fig.~\ref{fig:model_comparison_scatter}). As expected, both models assign lower scores for humans and higher for bots. High correlation coefficients indicate agreement between the models. 

\subsection{Feature Importance Analysis}

To compare the usefulness of different features, we trained models using each class of features alone. 
We achieved the best performance with user meta-data features; content features are also effective. Both yielded AUC above 0.9. Other feature classes yielded AUC above 0.8.

We analyzed the importance of single features using 
the Gini impurity score produced by our Random Forests model.
To rank the top features for a given dataset, we randomly select a subset of 10,000 accounts and compute the top features across 100 randomized experiments. 
The top 10 features 
are sufficient to reach performance of 0.9 AUC.
Sentiment and content of mentioned tweets are important features along with the statistical properties of retweet networks. Features of the friends with whom a user interacts are strong predictors as well. We observed the redundancy among many correlated features, such as distribution-type features (cf.~Table~\ref{tab:features}), especially in the content and sentiment categories. 
Further analysis of feature importance is the subject of ongoing investigation.

\subsection{False Positive and False Negative Cases}

Neither human annotators nor machine-learning models perform flawlessly. Humans are better at generalizing and learning new features from observed data. Machines outperform human annotators at processing large numbers of relations and searching for complex patterns. 
We analyzed our annotated accounts and their bot scores to highlight when disagreement occurs between annotators and classification models. Using an optimal threshold, we measured false positive and false negative rates at 0.15 and 0.11 respectively in our extended dataset. In these experiments, human annotation is considered as ground truth.

We identified the cases when the disagreement between classifier score and annotations occurs. We manually examined a sample from these accounts to investigate these errors. Accounts annotated as human can be classified as bot when an account posts tweets created by connected applications from other platforms. Some unusually active users are also classified as bots. Those users tend to have more retweets in general. This is somewhat intuitive as retweeting has lower cost than creating new content.
We encountered examples of misclassification for organizational and promotional accounts. Such accounts are often operated by multiple individuals, or combinations of users and automatic tools, generating misleading cues for the classifiers. Finally, the language of the content can also cause errors: our models tend to assign high bot scores to users who tweet in multiple languages. To mitigate this problem, the public version of our system now includes a classifier that ignores language-dependent features. 

\section{Estimation of Bot Population}
\label{sec:estimation}

\begin{figure*}[!th]
	\centering
	\includegraphics[width=0.24\linewidth]{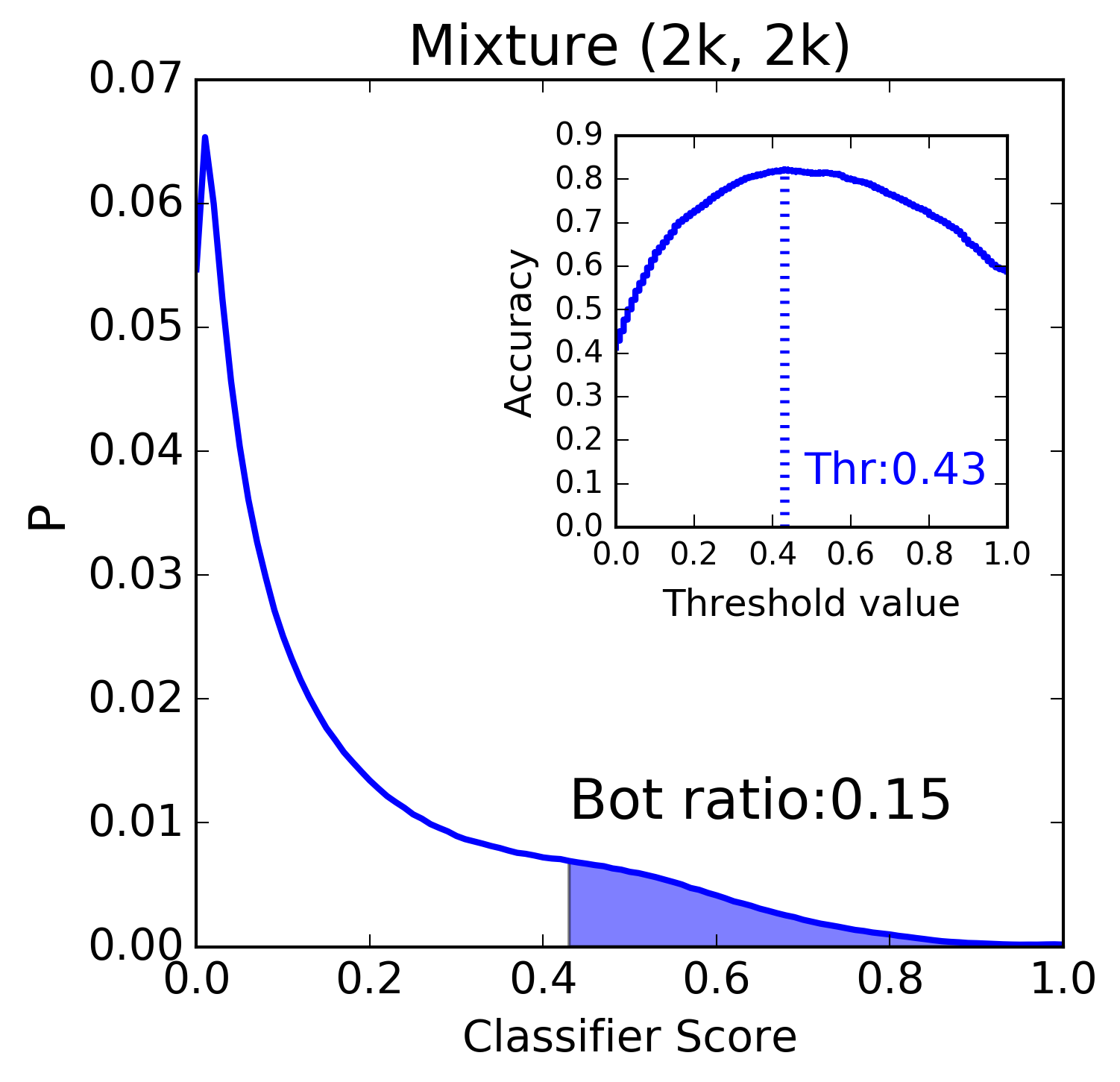}
	\includegraphics[width=0.24\linewidth]{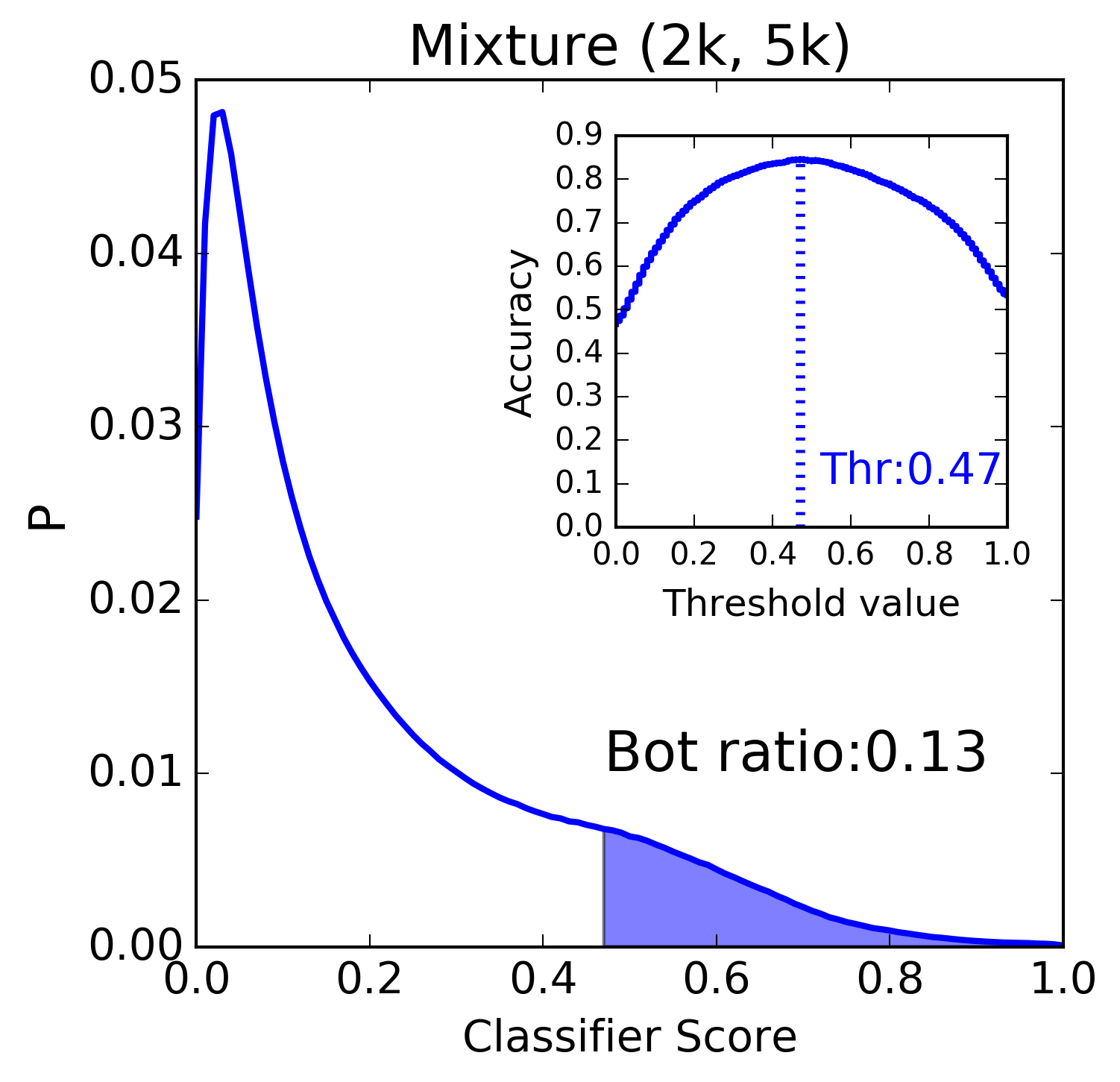}
	\includegraphics[width=0.24\linewidth]{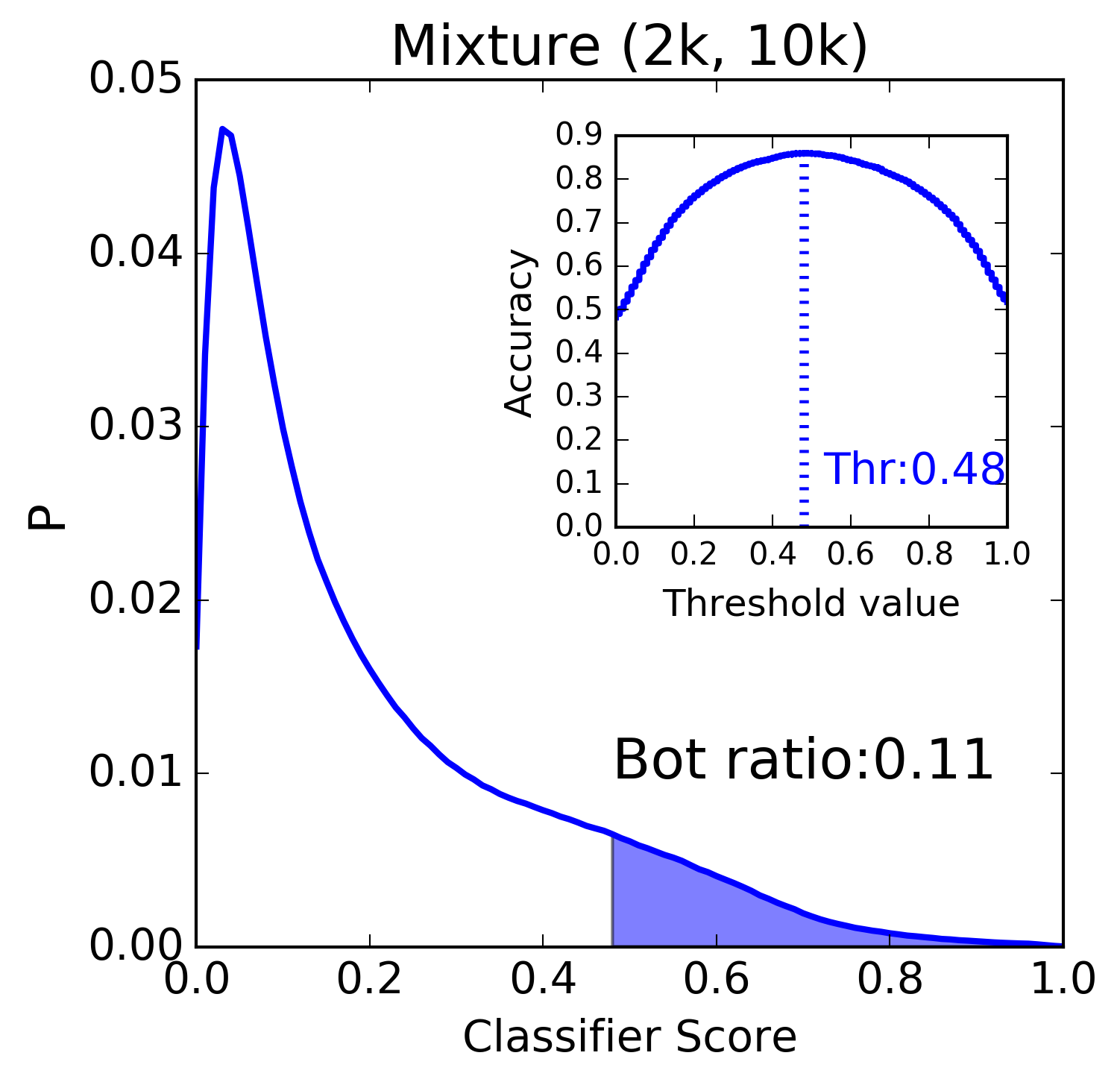}
	\includegraphics[width=0.24\linewidth]{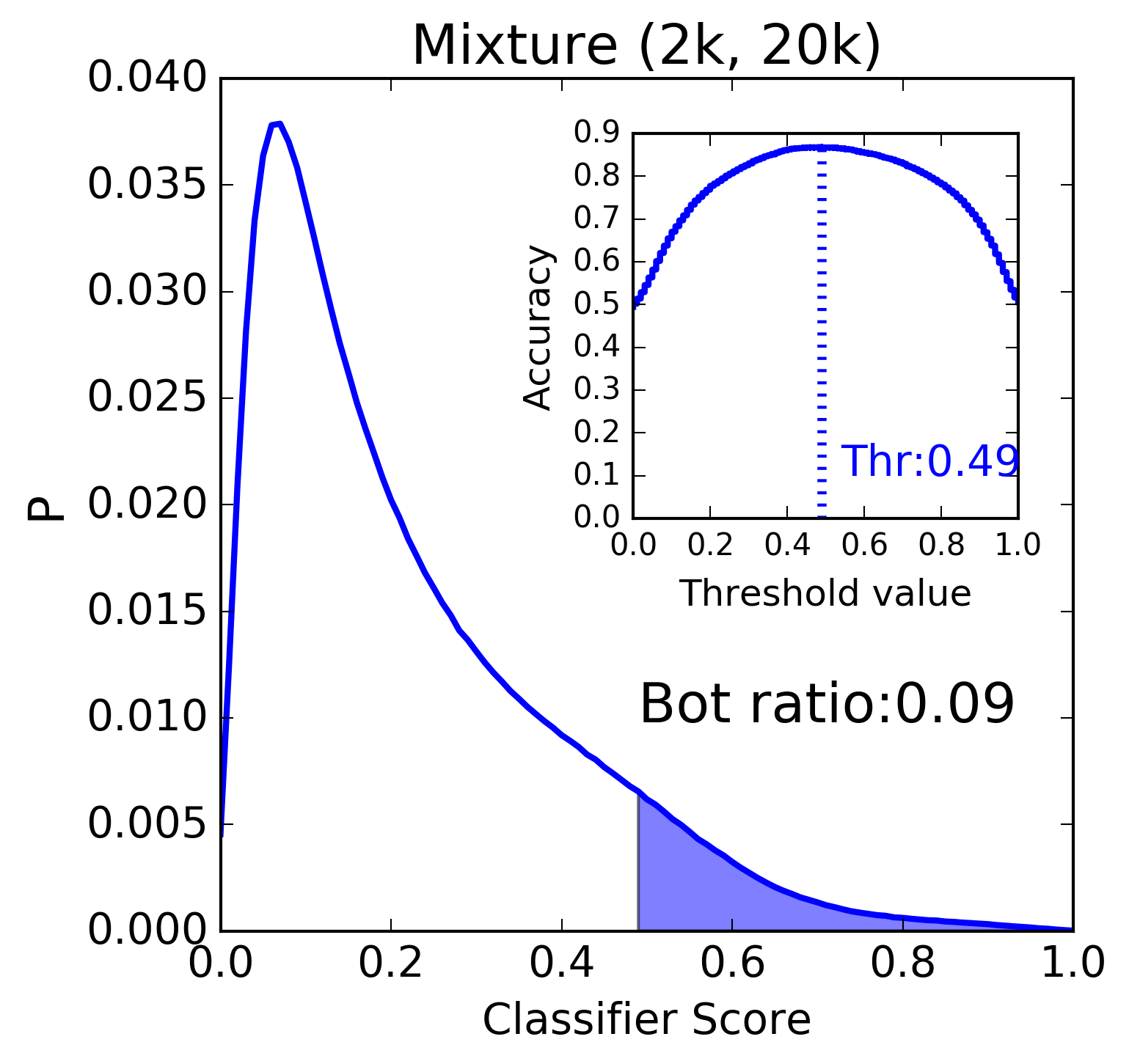}
	\caption{Estimation of bot population obtained from models with different sensitivity to sophisticated bots. The main charts show the score distributions based on our dataset of 14M users; accounts identified as bots are highlighted. The inset plots show how the thresholds are computed by maximizing accuracy. The titles of each subplot reflect the number of accounts from the annotated and honeypot datasets, respectively.}
	\label{fig:bot_estimations}
\end{figure*}

In a 2014 report by Twitter to the US Securities and Exchange Commission, the company put forth an estimate that between 5\% and 8.5\% of their user base consists of bots.\footnote{\url{www.sec.gov/Archives/edgar/data/1418091/000156459014003474/twtr-10q_20140630.htm}} We would like to offer our own assessment of the proportion of bot accounts as measured with our approach. Since our framework provides a continuous bot score as opposed to a discrete bot/human judgement, we must first determine an appropriate bot-score threshold separating human and bot accounts to estimate the proportion of bot accounts.

To infer a suitable threshold, we computed classification accuracies for varying thresholds considering all accounts scoring below each threshold as human. We then selected the threshold yielding maximum accuracy (see insets of Fig.~\ref{fig:bot_estimations}).

We estimated the population of bots using different models. This approach allows us to identify lower and upper bounds for the prevalence of Twitter bots. 
Models trained using the annotated dataset alone yield estimates of up to 15\% of accounts being bots.
Recall that the honeypot dataset was obtained earlier and therefore does not include newer, more sophisticated bots.
Thus models trained on the honeypot data alone are less sensitive to these sophisticated bots, yielding a more conservative estimate of 9\%.
Mixing the training data from these two sources results in estimates between these bounds depending on the ratio of the mixture, as illustrated in Fig.~\ref{fig:bot_estimations}.
Taken together, these numbers suggest that estimates about the prevalence of Twitter bots are highly dependent on the definition and sophistication of the  bots. 

Some other remarks are in order. First, we do not exclude the possibility that very sophisticated bots can systematically escape a human annotator's judgement. These complex bots may be active on Twitter, and therefore present in our datasets, and may have been incorrectly labeled as humans, making even the 15\% figure a conservative estimate.
Second, increasing  evidence suggests the presence on social media of hybrid human-bot accounts (sometimes referred to as \textit{cyborgs}) that perform automated actions with some human supervision~\cite{chu2012detecting,clark2016sifting}. Some have been allegedly used for terrorist propaganda and recruitment purposes. It remains unclear how these accounts should be labeled, and how pervasive they are.

\section{Characterization of User Interactions}

Let us next characterize social connectivity, information flow, and shared properties of users. We analyze the creation of social ties by accounts with different bot scores, and their interactions through shared content. We also cluster accounts and investigate shared properties of users in each cluster. Here and in the remainder of this paper, bot scores are computed with a model trained on the \emph{merged} dataset.

\subsection{Social connectivity}

\begin{figure}[!t]
	\centering
	\includegraphics[width=0.8\columnwidth]{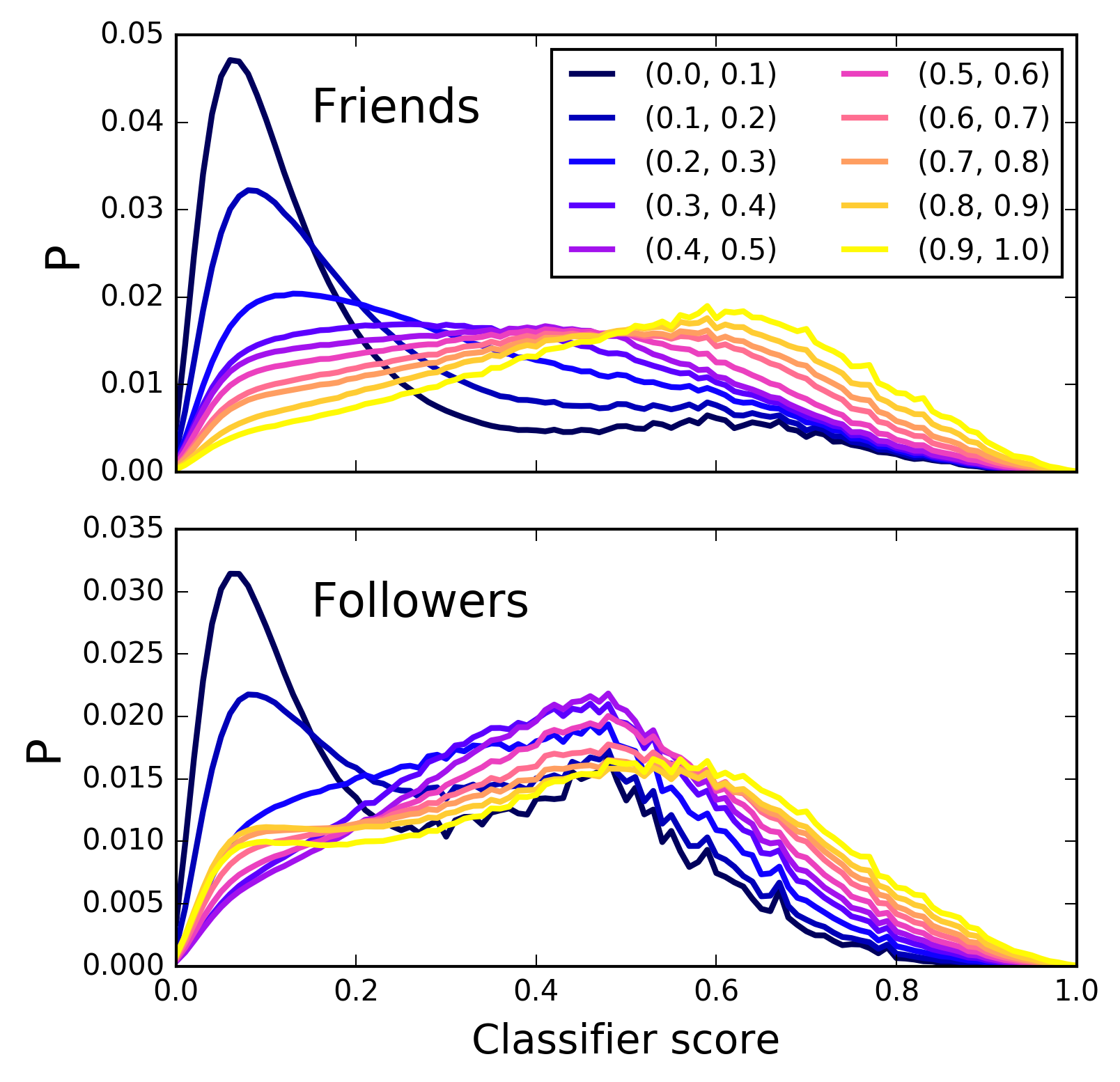}
	\caption{Distributions of bot scores for friends (top) and followers (bottom) of accounts in different score intervals.}
	\label{fig:social_connectivity}
\end{figure}

\begin{figure}[!t]
	\centering
    \includegraphics[width=0.8\columnwidth]{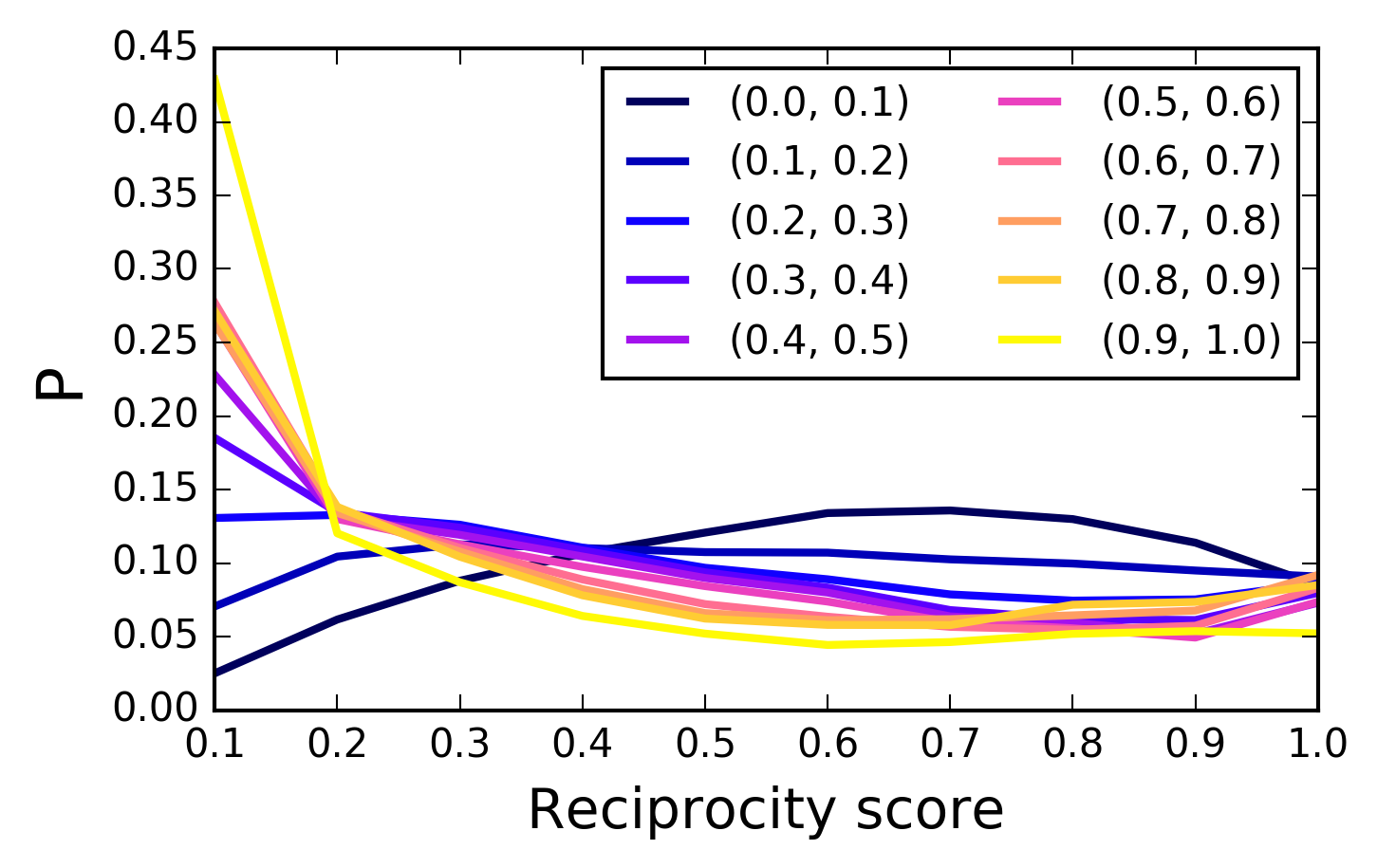}
    \caption{Distribution of reciprocity scores for accounts in different score intervals.}
    \label{fig:social_connectivity_reciprocal}
\end{figure}

To characterize the social connectivity, we collected the social networks of the accounts in our dataset using the Twitter API. Resulting friend and follower relations account for 46 billion social ties, 7 billion of which represent ties between the initially collected user set.

Our observations on social connectivity are presented in Fig.~\ref{fig:social_connectivity}. We computed bot-score distributions of friends and followers of accounts for each score interval.
The dark line in the top panel shows that human accounts (low score) mostly follow other human accounts. 
The dark line in the bottom panel shows a principal peak around 0.1 and a secondary one around 0.5. This indicates that humans are typically followed by other humans, but also by sophisticated bots (intermediate scores).
The lines corresponding to high scores in the two panels show 
that bots tend to follow other bots and they are mostly followed by bots.
However simple bots (0.8--1.0 ranges)  can also attract human attention.  
This happens when, e.g., humans follow benign bots such as those that share news. This gives rise to the secondary peak of the red line in the bottom panel.
In summary, the creation of social ties leads to a homophily effect. 

Fig.~\ref{fig:social_connectivity_reciprocal} illustrates the extent to which connections are reciprocated, given the nature of the accounts forming the ties. The \emph{reciprocity score} of a user is defined as the fraction of friends who are also followers. 
We observe that human accounts reciprocate more (dark line). Increasing bot scores correlate with lower reciprocity. We also observe that simple bot accounts (0.8--1.0 ranges) have  bimodal reciprocity distributions, indicating the existence of two distinct behaviors. The majority of high-score accounts have reciprocity score smaller than 0.2, possibly because simple bots follow users at random. 
The slight increase as the reciprocity score approaches one may be due to botnet accounts that coordinate by following each other.

\subsection{Information flow}

\begin{figure}[!t]
	\centering
	\includegraphics[width=0.8\columnwidth]{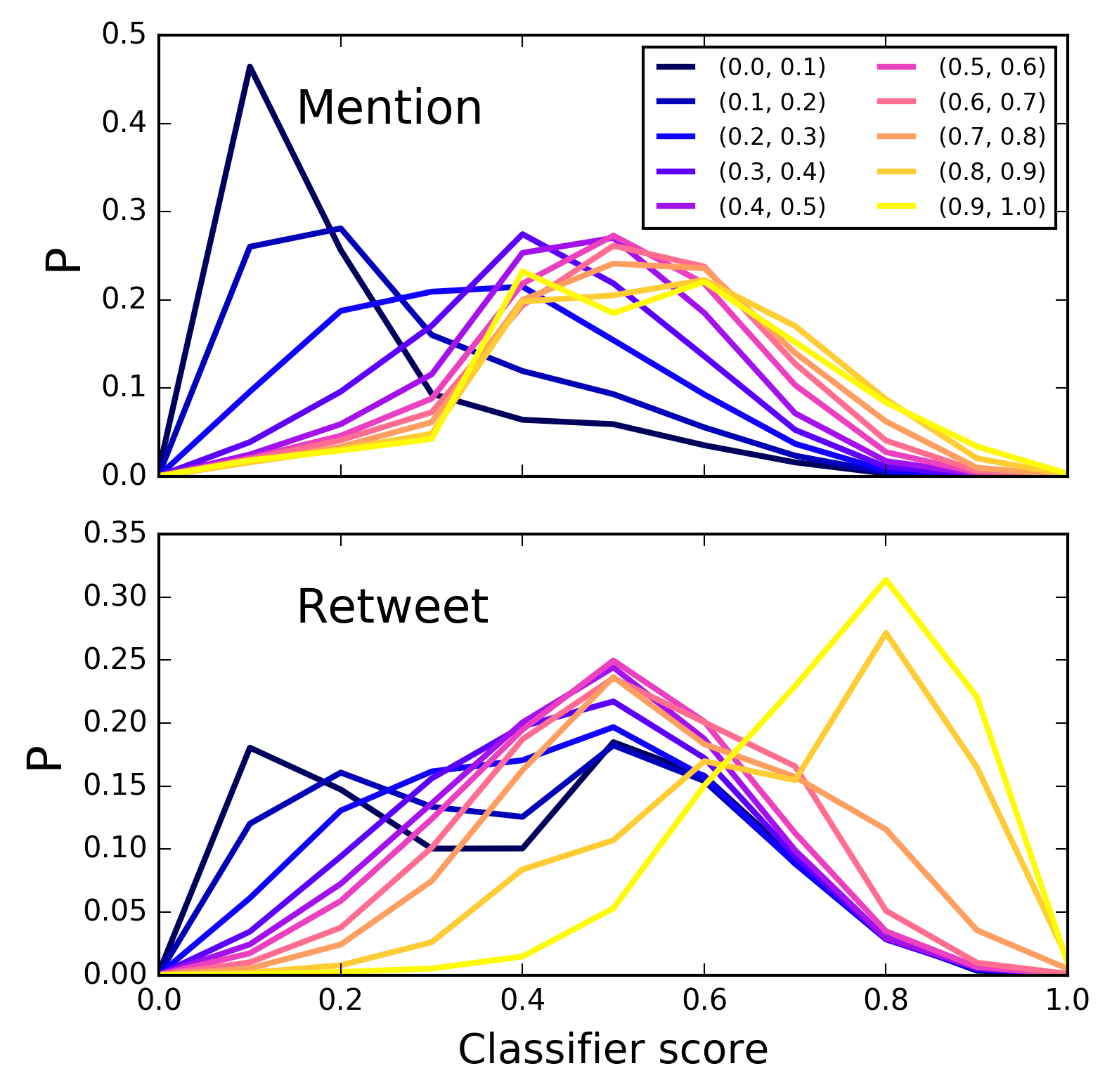}
	\caption{Bot score distributions of users mentioned (top) and retweeted (bottom) by accounts with different scores.}
	\label{fig:information_flow}
\end{figure}

Twitter is a platform that fosters social connectivity and the broadcasting of popular content. 
In Fig.~\ref{fig:information_flow} we analyze information flow in terms of mentions/retweets as a function of the score of the account being mentioned or retweeted.

Simple bots tend to retweet each other (lines for scores in the 0.8--1.0 ranges  peak around 0.8 in the bottom panel), while they frequently mention sophisticated bots (peaking around 0.5 in the top panel). 
More sophisticated bots (scores in the 0.5--0.7 ranges) retweet, but do not mention humans. They might be unable to engage in meaningful exchanges with humans.
While humans also retweet bots, as they may post interesting content (see peaks of the dark lines in the bottom panel), they have no interest in mentioning bots directly (dark lines in the top panel).

\subsection{Clustering accounts}

\begin{figure}[!t]
	\centering
	\includegraphics[width=\columnwidth]{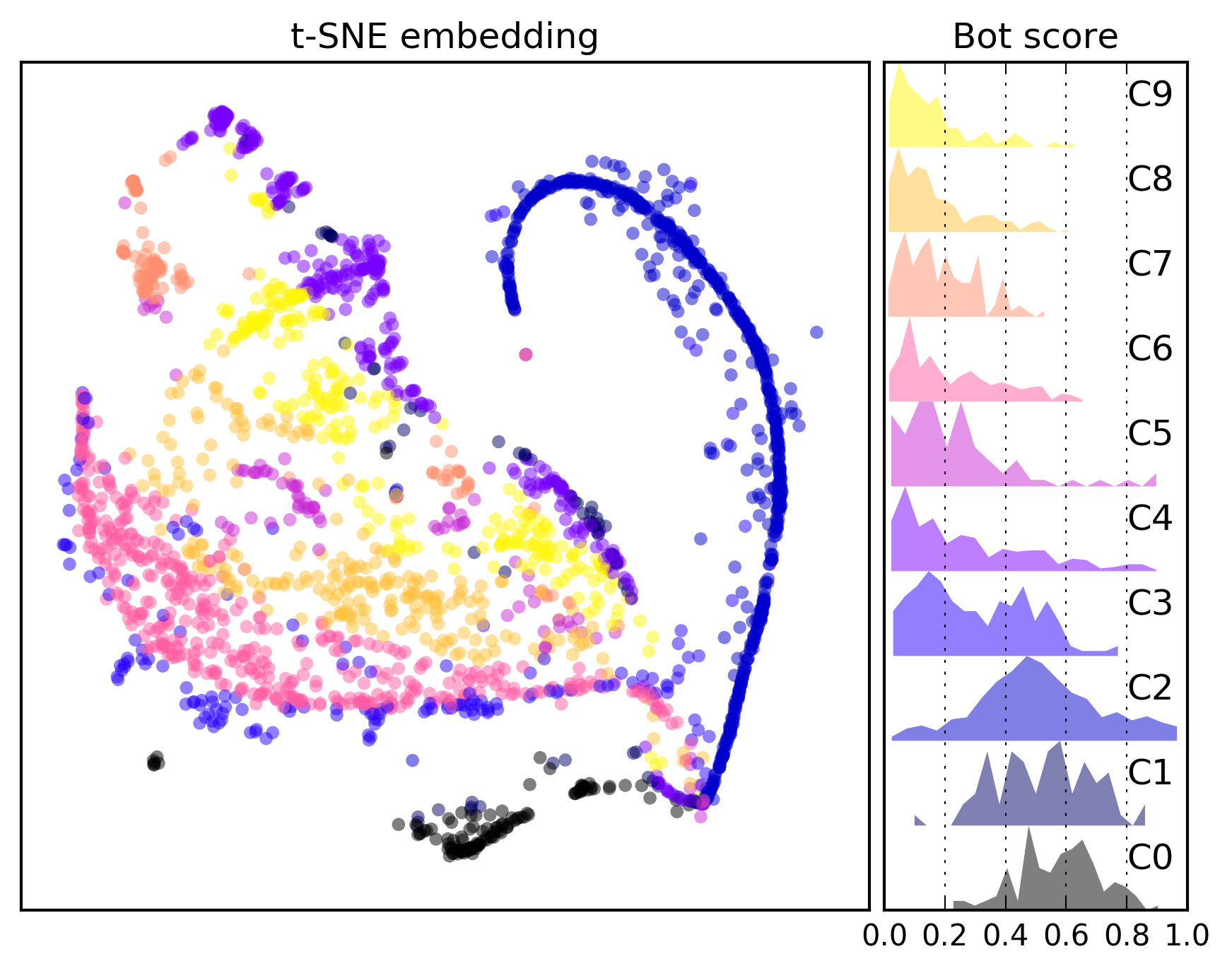}
	\caption{t-SNE embedding of accounts. Points are colored based on clustering in high-dimensional space. For each cluster, the distribution of scores is presented on the right.}
	\label{fig:account_clustering}
\end{figure}

To characterize different account types, let us group accounts into behavioral clusters. We apply K-Means to normalized vectors of the 100 most important features selected by our Random Forests model.
We identify 10 distinct clusters based on different evaluation criteria, such as silhouette scores and percentage of variance explained.
In Fig~\ref{fig:account_clustering}, we present a 2-dimensional projection of users  obtained by a dimensionality reduction technique called t-SNE~\cite{maaten2008visualizing}. 
In this method, the similarity between users is computed based on their 100-dimensional representation in the feature space. Similar users are  projected into nearby points and dissimilar users are kept distant from each other.

Let us investigate shared cluster properties  by manual inspection of random subsets of accounts from each cluster. 
Three of the clusters, namely \cluster{0}--\cluster{2}, have high average bot scores. The presence of significant amounts of bot accounts in these clusters was manually verified. 
These \textit{bot} clusters exhibit some prominent properties: cluster \cluster{0}, for example, consists of legit-looking accounts that are promoting themselves (recruiters, porn actresses, etc.). They are concentrated in the lower part of the 2-dimensional embedding, suggesting homogeneous patterns of behaviors. \cluster{1} contains spam accounts that are very active but have few followers. Accounts in \cluster{2} frequently use automated applications to share activity from other platforms like YouTube and Instagram, or post links to news articles. Some of the accounts in \cluster{2} might belong to actual humans who are no longer active and their posts are mostly sent by connected apps.

Cluster \cluster{3} contain a \textit{mix} of sophisticated bots, cyborg-like accounts (mix of bot and human features), and human users. 
Clusters of predominantly \textit{human} accounts, namely \cluster{4}--\cluster{9}, separate from one another in the embedding due to different activity styles, user popularity, content production and consumption patterns.
For instance, accounts in \cluster{7} engage more with their friends, unlike accounts from \cluster{8} that mostly retweet with little other forms of interaction. Clusters \cluster{5}, \cluster{6}, and \cluster{9} contain common Twitter users who produce experiential tweets, share pictures, and retweet their friends.

\section{Related Work}

Also known as ``sybil'' accounts, social bots can pollute online discussion by lending false credibility to their messages and influence other users~\cite{ferrara2014rise,aiello2014people}. Recent studies quantify the extent to which automated systems can dominate discussions on Twitter about topics ranging from electronic cigarettes~\cite{clark2015vaporous} to elections~\cite{bessi2016social}. 
Large collections of social bots, also known as botnets, are controlled by botmasters and used for coordinated activities. Examples of such botnets identified for advertisement~\cite{zhou2017starwars} and influence about Syrian civic war~\cite{abokhodair2015dissecting}.
Social bots also vary greatly in terms of their behavior, intent, and vulnerabilities, as illustrated in a categorization scheme for bot attacks~\cite{mitter2014categorization}.

Much of the previous work on detecting bots is from the perspective of the social network platform operators, implying full access to all data.
These studies focus on collecting large-scale data to either cluster behavioral patterns of users~\cite{wang2013you} or classify accounts using supervised learning techniques~\cite{yang2014uncovering,lee2011seven}. For instance, Beutel \textit{et al.} decomposed event data in time, user, and activity dimensions to extract similar behaviors~\cite{beutel2013copycatch}. These techniques are useful to identify coordinated large-scale attacks directed at a common set of targets at the same time, but accounts with similar strategies might also target different groups and operate separately from each other. 

Structural connectivity may provide important cues. However, Yang \textit{et al.} studied large-scale sybil attacks and observed sophisticated sybils that develop strategies for building normal-looking social ties, making themselves harder to detect~\cite{yang2014uncovering}. Some sybil attacks analyze the social graph of targeted groups to infiltrate specific organizations~\cite{elyashar2013homing}. SybilRank is a system developed to identify attacks from their underlying topology~\cite{cao2012aiding}.
Alvisi \textit{et al.} surveyed the evolution of sybil defense protocols that leverage the structural properties of the social graph~\cite{alvisi2013sok}. 

The work presented here follows several previous contributions to the problem of social bot detection that leverage learning models trained with data collected from human and  bot accounts. Chu \textit{et al.} built a classification system identifying accounts controlled by humans, bots, and cyborgs~\cite{chu2010tweeting,chu2012detecting}. Wang \textit{et al.} analyzed sybil attacks using annotations by experts and crowd-sourcing workers to evaluate consistency and effectiveness of different detection systems~\cite{wang2012social}. 
Clark \textit{et al.} labeled 1,000 accounts by hand and found natural language text features to be very effective at discriminating between human and automated accounts~\cite{clark2016sifting}. Lee \textit{et al.}  used a honeypot approach to collect the largest sample of bot accounts available to date~\cite{lee2011seven}. That study generated the honeypot dataset used in the present paper. Here, we extend this body of prior work by exploring many different categories of features, contributing a new labeled dataset, estimating the number of bot accounts, analyzing information flow among accounts, identifying several classes of behaviors, and providing a public bot detection service. 

An alternative approach to study social bots and sybil attacks is to understand what makes certain groups and individuals more appealing as targets. Wald \textit{et al.} studied the factors affecting the likelihood of a users being targeted by social bots~\cite{wald2013predicting}. 
These approaches point to effective strategies that future social bots might develop. 

Recently, we have observed efforts to facilitate research collaborations on the topic of social bots. DARPA organized a bot detection challenge in the domain of anti-vaccine campaigns on Twitter~\cite{subrahmanian2016darpa}. 
We released our Twitter bot detection system online for public use~\cite{davis2016botornot}. Since its release, our system has received millions of requests and we are improving models based on feedback we received from our users.
The increasing availability of software and datasets on social bots will help design systems that are capable of co-evolving with recent social bots and hopefully mitigating the effects of their malicious activities.

\section{Conclusions}

Social media make it easy for accounts controlled by hybrid or automated approaches to create content and interact with other accounts. Our project aims to identify these bots. Such a classification task could be a first step toward studying modes of communication among different classes of entities on social media.

In this article, we presented a framework for bot detection on Twitter. We introduced our machine learning system that extracts more than a thousand features in six different classes: users and friends meta-data, tweet content and sentiment, network patterns, and activity time series. We evaluated our framework when initially trained on an available dataset of bots. Our initial classifier achieves 0.95 AUC when evaluated by using 5-fold cross validation. Our analysis on the contributions of different feature classes suggests that user meta-data and content features are the two most valuable sources of data to detect simple bots. 

To evaluate the performance of our classifier on a more recent and challenging sample of bots, we randomly selected Twitter accounts covering the whole spectrum of classification scores. The accuracy of our initial classifier trained on the honeypot dataset decreased to 0.85 AUC when tested on the more challenging dataset. By retraining the classifier with the two datasets merged, we achieved high accuracy (0.94 AUC) in detecting both simple and sophisticated bots.

We also estimated the fraction of bots in the active English-speaking population on Twitter. We classified nearly 14M accounts using our system and inferred the optimal threshold scores that separate human and bot accounts for several models with different mixes of simple and sophisticated bots. Training data have an important effect on classifier sensitivity. Our estimates for the bot population range between 9\% and 15\%. This points to the importance of tracking increasingly sophisticated bots, since deception and detection technologies are in a never-ending arms race.

To characterize user interactions, we studied social connectivity and information flow between different user groups. We showed that selection of friends and followers are correlated with accounts bot-likelihood. We also highlighted how bots use different retweet and mention strategies when interacting with humans or other bots.

We concluded our analysis by characterizing subclasses of account behaviors. Clusters identified by this analysis point mainly to three types of bots. These results emphasize that Twitter hosts a variety of users with diverse behaviors; this is true for both human and bot accounts. In some cases, the boundary separating these two groups is not sharp and an account can exhibit characteristics of both.

\fontsize{9pt}{10pt}
\selectfont
\paragraph{Acknowledgments.}
We thank M. JafariAsbagh, P. Shiralkar for helpful discussions. We also want to thank undergraduate students A. Toms, A. Fulton, A. Witulski, and M. Johnston for contributing data annotation. This work was supported in part by ONR (N15A-020-0053), DARPA (W911NF-12-1-0037), NSF (CCF-1101743), and the J.S. McDonnell Foundation. 

%

\fontsize{9pt}{10pt}
\selectfont
\bibliographystyle{aaai}
\bibliography{sigproc}  

\begin{thebibliography}{}

\bibitem[\protect\citeauthoryear{Abokhodair, Yoo, and
  McDonald}{2015}]{abokhodair2015dissecting}
Abokhodair, N.; Yoo, D.; and McDonald, D.~W.
\newblock 2015.
\newblock Dissecting a social botnet: Growth, content and influence in twitter.
\newblock In {\em Proc. of the 18th ACM Conf. on Computer Supported Cooperative
  Work \& Social Computing},  839--851.
\newblock ACM.

\bibitem[\protect\citeauthoryear{Agarwal \bgroup et al\mbox.\egroup
  }{2011}]{agarwal2011sentiment}
Agarwal, A.; Xie, B.; Vovsha, I.; Rambow, O.; and Passonneau, R.
\newblock 2011.
\newblock Sentiment analysis of {T}witter data.
\newblock In {\em Proc. of the Workshop on Languages in Social Media},  30--38.
\newblock ACL.

\bibitem[\protect\citeauthoryear{Aiello \bgroup et al\mbox.\egroup
  }{2012}]{aiello2014people}
Aiello, L.; Deplano, M.; Schifanella, R.; and Ruffo, G.
\newblock 2012.
\newblock People are strange when you're a stranger: Impact and influence of
  bots on social networks.
\newblock In {\em Proc. 6th Intl. AAAI Conf. on Weblogs \& Soc. Media (ICWSM)}.

\bibitem[\protect\citeauthoryear{Alvisi \bgroup et al\mbox.\egroup
  }{2013}]{alvisi2013sok}
Alvisi, L.; Clement, A.; Epasto, A.; Lattanzi, S.; and Panconesi, A.
\newblock 2013.
\newblock Sok: The evolution of sybil defense via social networks.
\newblock In {\em Proc. IEEE Symposium on Security and Privacy (SP)},
  382--396.

\bibitem[\protect\citeauthoryear{Bakshy \bgroup et al\mbox.\egroup
  }{2011}]{bakshy2011everyone}
Bakshy, E.; Hofman, J.~M.; Mason, W.~A.; and Watts, D.~J.
\newblock 2011.
\newblock Everyone's an influencer: quantifying influence on {T}witter.
\newblock In {\em Proc. 4th ACM Intl. Conf. on Web Search and Data Mining},
  65--74.

\bibitem[\protect\citeauthoryear{Berger and Morgan}{2015}]{berger2015isis}
Berger, J., and Morgan, J.
\newblock 2015.
\newblock The isis twitter census: Defining and describing the population of
  isis supporters on twitter.
\newblock {\em The Brookings Project on US Relations with the Islamic World}
  3:20.

\bibitem[\protect\citeauthoryear{Bessi and Ferrara}{2016}]{bessi2016social}
Bessi, A., and Ferrara, E.
\newblock 2016.
\newblock Social bots distort the 2016 us presidential election online
  discussion.
\newblock {\em First Monday} 21(11).

\bibitem[\protect\citeauthoryear{Bessi \bgroup et al\mbox.\egroup
  }{2015}]{bessi2015science}
Bessi, A.; Coletto, M.; Davidescu, G.~A.; Scala, A.; Caldarelli, G.; and
  Quattrociocchi, W.
\newblock 2015.
\newblock Science vs conspiracy: Collective narratives in the age of
  misinformation.
\newblock {\em PLoS ONE} 10(2):e0118093.

\bibitem[\protect\citeauthoryear{Beutel \bgroup et al\mbox.\egroup
  }{2013}]{beutel2013copycatch}
Beutel, A.; Xu, W.; Guruswami, V.; Palow, C.; and Faloutsos, C.
\newblock 2013.
\newblock Copycatch: stopping group attacks by spotting lockstep behavior in
  social networks.
\newblock In {\em Prov. 22nd Intl. ACM Conf. World Wide Web (WWW)},  119--130.

\bibitem[\protect\citeauthoryear{Bollen, Mao, and
  Zeng}{2011}]{bollen2011twitter}
Bollen, J.; Mao, H.; and Zeng, X.
\newblock 2011.
\newblock {T}witter mood predicts the stock market.
\newblock {\em Journal of Computational Science} 2(1):1--8.

\bibitem[\protect\citeauthoryear{Boshmaf \bgroup et al\mbox.\egroup
  }{2011}]{boshmaf2011socialbot}
Boshmaf, Y.; Muslukhov, I.; Beznosov, K.; and Ripeanu, M.
\newblock 2011.
\newblock The socialbot network: when bots socialize for fame and money.
\newblock In {\em Proc. 27th Annual Computer Security Applications Conf.}

\bibitem[\protect\citeauthoryear{Botta, Moat, and
  Preis}{2015}]{botta2015quantifying}
Botta, F.; Moat, H.~S.; and Preis, T.
\newblock 2015.
\newblock Quantifying crowd size with mobile phone and twitter data.
\newblock {\em Royal Society open science} 2(5):150162.

\bibitem[\protect\citeauthoryear{Briscoe, Appling, and
  Hayes}{2014}]{briscoe2014cues}
Briscoe, E.; Appling, S.; and Hayes, H.
\newblock 2014.
\newblock Cues to deception in social media communications.
\newblock In {\em Hawaii Intl. Conf. on Syst Sci}.

\bibitem[\protect\citeauthoryear{Cao \bgroup et al\mbox.\egroup
  }{2012}]{cao2012aiding}
Cao, Q.; Sirivianos, M.; Yang, X.; and Pregueiro, T.
\newblock 2012.
\newblock Aiding the detection of fake accounts in large scale social online
  services.
\newblock In {\em 9th USENIX Symp on Netw Sys Design \& Implement},  197--210.

\bibitem[\protect\citeauthoryear{Chavoshi, Hamooni, and
  Mueen}{2016}]{Chavoshi2016}
Chavoshi, N.; Hamooni, H.; and Mueen, A.
\newblock 2016.
\newblock Identifying correlated bots in twitter.
\newblock In {\em Social Informatics: 8th Intl. Conf.},  14--21.

\bibitem[\protect\citeauthoryear{Chu \bgroup et al\mbox.\egroup
  }{2010}]{chu2010tweeting}
Chu, Z.; Gianvecchio, S.; Wang, H.; and Jajodia, S.
\newblock 2010.
\newblock Who is tweeting on twitter: human, bot, or cyborg?
\newblock In {\em Proc. 26th annual computer security applications conf.},
  21--30.

\bibitem[\protect\citeauthoryear{Chu \bgroup et al\mbox.\egroup
  }{2012}]{chu2012detecting}
Chu, Z.; Gianvecchio, S.; Wang, H.; and Jajodia, S.
\newblock 2012.
\newblock Detecting automation of twitter accounts: Are you a human, bot, or
  cyborg?
\newblock {\em IEEE Tran Dependable \& Secure Comput} 9(6):811--824.

\bibitem[\protect\citeauthoryear{Clark \bgroup et al\mbox.\egroup
  }{2015}]{clark2015vaporous}
Clark, E.; Jones, C.; Williams, J.; Kurti, A.; Nortotsky, M.; Danforth, C.; and
  Dodds, P.
\newblock 2015.
\newblock Vaporous marketing: Uncovering pervasive electronic cigarette
  advertisements on twitter.
\newblock {\em arXiv preprint arXiv:1508.01843}.

\bibitem[\protect\citeauthoryear{Clark \bgroup et al\mbox.\egroup
  }{2016}]{clark2016sifting}
Clark, E.; Williams, J.; Jones, C.; Galbraith, R.; Danforth, C.; and Dodds, P.
\newblock 2016.
\newblock Sifting robotic from organic text: a natural language approach for
  detecting automation on twitter.
\newblock {\em Journal of Computational Science} 16:1--7.

\bibitem[\protect\citeauthoryear{Danescu-Niculescu-Mizil \bgroup et
  al\mbox.\egroup }{2013}]{danescu2013no}
Danescu-Niculescu-Mizil, C.; West, R.; Jurafsky, D.; Leskovec, J.; and Potts,
  C.
\newblock 2013.
\newblock No country for old members: user lifecycle and linguistic change in
  online communities.
\newblock In {\em Proc. of the 22nd Intl. Conf. on World Wide Web},  307--318.

\bibitem[\protect\citeauthoryear{Das \bgroup et al\mbox.\egroup
  }{2016}]{hin2016}
Das, A.; Gollapudi, S.; Kiciman, E.; and Varol, O.
\newblock 2016.
\newblock Information dissemination in heterogeneous-intent networks.
\newblock In {\em Proc. ACM Conf. on Web Science}.

\bibitem[\protect\citeauthoryear{Davis \bgroup et al\mbox.\egroup
  }{2016}]{davis2016botornot}
Davis, C.~A.; Varol, O.; Ferrara, E.; Flammini, A.; and Menczer, F.
\newblock 2016.
\newblock {BotOrNot: A system to evaluate social bots}.
\newblock In {\em Proc. 25th Intl. Conf. Companion on World Wide Web},
  273--274.

\bibitem[\protect\citeauthoryear{Echeverr{\'\i}a and
  Zhou}{2017}]{zhou2017starwars}
Echeverr{\'\i}a, J., and Zhou, S.
\newblock 2017.
\newblock The `star wars' botnet with >350k twitter bots.
\newblock {\em arXiv preprint arXiv:1701.02405}.

\bibitem[\protect\citeauthoryear{Elyashar \bgroup et al\mbox.\egroup
  }{2013}]{elyashar2013homing}
Elyashar, A.; Fire, M.; Kagan, D.; and Elovici, Y.
\newblock 2013.
\newblock Homing socialbots: intrusion on a specific organization's employee
  using socialbots.
\newblock In {\em Proc. IEEE/ACM Intl. Conf. on Advances in Social Networks
  Analysis and Mining},  1358--1365.

\bibitem[\protect\citeauthoryear{Ferrara and
  Yang}{2015}]{ferrara2015quantifying}
Ferrara, E., and Yang, Z.
\newblock 2015.
\newblock Quantifying the effect of sentiment on information diffusion in
  social media.
\newblock {\em PeerJ Comp. Sci.} 1:e26.

\bibitem[\protect\citeauthoryear{Ferrara \bgroup et al\mbox.\egroup
  }{2016a}]{ferrara2014rise}
Ferrara, E.; Varol, O.; Davis, C.; Menczer, F.; and Flammini, A.
\newblock 2016a.
\newblock The rise of social bots.
\newblock {\em Comm. ACM} 59(7):96--104.

\bibitem[\protect\citeauthoryear{Ferrara \bgroup et al\mbox.\egroup
  }{2016b}]{ferrara2016campaign}
Ferrara, E.; Varol, O.; Menczer, F.; and Flammini, A.
\newblock 2016b.
\newblock Detection of promoted social media campaigns.
\newblock In {\em Proc. Intl. AAAI Conference on Web and Social Media}.

\bibitem[\protect\citeauthoryear{Ferrara \bgroup et al\mbox.\egroup
  }{2016c}]{ferrara2016predicting}
Ferrara, E.; Wang, W.-Q.; Varol, O.; Flammini, A.; and Galstyan, A.
\newblock 2016c.
\newblock Predicting online extremism, content adopters, and interaction
  reciprocity.
\newblock In {\em Social Informatics: 8th Intl. Conf., SocInfo 2016, Bellevue,
  WA, USA},  22--39.

\bibitem[\protect\citeauthoryear{Ghosh, Surachawala, and
  Lerman}{2011}]{Ghosh11snakdd}
Ghosh, R.; Surachawala, T.; and Lerman, K.
\newblock 2011.
\newblock Entropy-based classification of retweeting activity on twitter.
\newblock In {\em Proc. of KDD workshop on Social Network Analysis}.

\bibitem[\protect\citeauthoryear{Gjoka \bgroup et al\mbox.\egroup
  }{2010}]{gjoka2010walking}
Gjoka, M.; Kurant, M.; Butts, C.~T.; and Markopoulou, A.
\newblock 2010.
\newblock Walking in facebook: A case study of unbiased sampling of osns.
\newblock In {\em Proc. IEEE INFOCOM},  1--9.

\bibitem[\protect\citeauthoryear{Haustein \bgroup et al\mbox.\egroup
  }{2016}]{haustein2016tweets}
Haustein, S.; Bowman, T.~D.; Holmberg, K.; Tsou, A.; Sugimoto, C.~R.; and
  Larivi{\`e}re, V.
\newblock 2016.
\newblock Tweets as impact indicators: Examining the implications of automated
  ``bot'' accounts on twitter.
\newblock {\em Journal of the Association for Information Science and
  Technology} 67(1):232--238.

\bibitem[\protect\citeauthoryear{Kloumann \bgroup et al\mbox.\egroup
  }{2012}]{kloumann2012positivity}
Kloumann, I.~M.; Danforth, C.~M.; Harris, K.~D.; Bliss, C.~A.; and Dodds, P.~S.
\newblock 2012.
\newblock Positivity of the english language.
\newblock {\em PLoS ONE} 7(1):e29484.

\bibitem[\protect\citeauthoryear{Lee, Eoff, and Caverlee}{2011}]{lee2011seven}
Lee, K.; Eoff, B.~D.; and Caverlee, J.
\newblock 2011.
\newblock Seven months with the devils: A long-term study of content polluters
  on twitter.
\newblock In {\em Proc. 5th AAAI Intl. Conf. on Web and Social Media}.

\bibitem[\protect\citeauthoryear{Letchford, Moat, and
  Preis}{2015}]{letchford2015advantage}
Letchford, A.; Moat, H.~S.; and Preis, T.
\newblock 2015.
\newblock The advantage of short paper titles.
\newblock {\em Royal Society Open Science} 2(8):150266.

\bibitem[\protect\citeauthoryear{Lokot and Diakopoulos}{2016}]{lokot2016news}
Lokot, T., and Diakopoulos, N.
\newblock 2016.
\newblock News bots: Automating news and information dissemination on twitter.
\newblock {\em Digital Journalism} 4(6):682--699.

\bibitem[\protect\citeauthoryear{Maaten and
  Hinton}{2008}]{maaten2008visualizing}
Maaten, L. v.~d., and Hinton, G.
\newblock 2008.
\newblock Visualizing data using t-sne.
\newblock {\em Journal of Machine Learning Research} 9(Nov):2579--2605.

\bibitem[\protect\citeauthoryear{McAuley and
  Leskovec}{2013}]{mcauley2013amateurs}
McAuley, J., and Leskovec, J.
\newblock 2013.
\newblock From amateurs to connoisseurs: modeling the evolution of user
  expertise through online reviews.
\newblock In {\em Proc. 22nd Intl. ACM Conf. World Wide Web},  897--908.

\bibitem[\protect\citeauthoryear{Mislove \bgroup et al\mbox.\egroup
  }{2011}]{mislove2011understanding}
Mislove, A.; Lehmann, S.; Ahn, Y.-Y.; Onnela, J.-P.; and Rosenquist, J.~N.
\newblock 2011.
\newblock Understanding the demographics of {T}witter users.
\newblock In {\em Proc. of the 5th Intl. AAAI Conf. on Weblogs and Social
  Media}.

\bibitem[\protect\citeauthoryear{Mitchell \bgroup et al\mbox.\egroup
  }{2013}]{mitchell2013geography}
Mitchell, L.; Harris, K.~D.; Frank, M.~R.; Dodds, P.~S.; and Danforth, C.~M.
\newblock 2013.
\newblock The geography of happiness: Connecting {T}witter sentiment and
  expression, demographics, and objective characteristics of place.
\newblock {\em PLoS ONE} 8(5):e64417.

\bibitem[\protect\citeauthoryear{Mitter, Wagner, and
  Strohmaier}{2013}]{mitter2014categorization}
Mitter, S.; Wagner, C.; and Strohmaier, M.
\newblock 2013.
\newblock A categorization scheme for socialbot attacks in online social
  networks.
\newblock In {\em Proc. of the 3rd ACM Web Science Conference}.

\bibitem[\protect\citeauthoryear{Mocanu \bgroup et al\mbox.\egroup
  }{2013}]{mocanu2013twitter}
Mocanu, D.; Baronchelli, A.; Perra, N.; Gon\c{c}alves, B.; Zhang, Q.; and
  Vespignani, A.
\newblock 2013.
\newblock The {T}witter of {B}abel: Mapping world languages through
  microblogging platforms.
\newblock {\em PLoS ONE} 8(4):e61981.

\bibitem[\protect\citeauthoryear{Morstatter \bgroup et al\mbox.\egroup
  }{2013}]{morstatter2013sample}
Morstatter, F.; Pfeffer, J.; Liu, H.; and Carley, K.
\newblock 2013.
\newblock Is the sample good enough? comparing data from twitter's streaming
  api with twitter's firehose.
\newblock In {\em 7th Int Conf on Weblogs \& Soc Med}.

\bibitem[\protect\citeauthoryear{Pedregosa \bgroup et al\mbox.\egroup
  }{2011}]{scikit-learn}
Pedregosa, F.; Varoquaux, G.; Gramfort, A.; Michel, V.; Thirion, B.; Grisel,
  O.; et~al.
\newblock 2011.
\newblock Scikit-learn: Machine learning in {P}ython.
\newblock {\em Journal of Machine Learning Research} 12:2825--2830.

\bibitem[\protect\citeauthoryear{Ratkiewicz \bgroup et al\mbox.\egroup
  }{2011}]{ratkiewicz2011detecting}
Ratkiewicz, J.; Conover, M.; Meiss, M.; Goncalves, B.; Flammini, A.; and
  Menczer, F.
\newblock 2011.
\newblock Detecting and tracking political abuse in social media.
\newblock In {\em 5th Int Conf on Weblogs \& Soc Med},  297--304.

\bibitem[\protect\citeauthoryear{Savage, Monroy-Hernandez, and
  H{\"o}llerer}{2016}]{savage2016botivist}
Savage, S.; Monroy-Hernandez, A.; and H{\"o}llerer, T.
\newblock 2016.
\newblock Botivist: Calling volunteers to action using online bots.
\newblock In {\em Proceedings of the 19th ACM Conference on Computer-Supported
  Cooperative Work \& Social Computing},  813--822.
\newblock ACM.

\bibitem[\protect\citeauthoryear{Subrahmanian \bgroup et al\mbox.\egroup
  }{2016}]{subrahmanian2016darpa}
Subrahmanian, V.; Azaria, A.; Durst, S.; Kagan, V.; Galstyan, A.; Lerman, K.;
  Zhu, L.; Ferrara, E.; Flammini, A.; Menczer, F.; et~al.
\newblock 2016.
\newblock {The DARPA Twitter Bot Challenge}.
\newblock {\em IEEE Computer} 6(49):38--46.

\bibitem[\protect\citeauthoryear{Wald \bgroup et al\mbox.\egroup
  }{2013}]{wald2013predicting}
Wald, R.; Khoshgoftaar, T.~M.; Napolitano, A.; and Sumner, C.
\newblock 2013.
\newblock Predicting susceptibility to social bots on twitter.
\newblock In {\em Proc. 14th Intl. IEEE Conf. on Information Reuse and
  Integration},  6--13.

\bibitem[\protect\citeauthoryear{Wang \bgroup et al\mbox.\egroup
  }{2013a}]{wang2013you}
Wang, G.; Konolige, T.; Wilson, C.; Wang, X.; Zheng, H.; and Zhao, B.~Y.
\newblock 2013a.
\newblock You are how you click: Clickstream analysis for sybil detection.
\newblock In {\em Proc. USENIX Security},  1--15.
\newblock Citeseer.

\bibitem[\protect\citeauthoryear{Wang \bgroup et al\mbox.\egroup
  }{2013b}]{wang2012social}
Wang, G.; Mohanlal, M.; Wilson, C.; Wang, X.; Metzger, M.; Zheng, H.; and Zhao,
  B.~Y.
\newblock 2013b.
\newblock Social turing tests: Crowdsourcing sybil detection.
\newblock In {\em Proc. of the 20th Network \& Distributed System Security
  Symposium (NDSS)}.

\bibitem[\protect\citeauthoryear{Warriner, Kuperman, and
  Brysbaert}{2013}]{warriner2013norms}
Warriner, A.~B.; Kuperman, V.; and Brysbaert, M.
\newblock 2013.
\newblock Norms of valence, arousal, and dominance for 13,915 english lemmas.
\newblock {\em Behavior research methods}  1--17.

\bibitem[\protect\citeauthoryear{Wilson, Wiebe, and
  Hoffmann}{2005}]{wilson2005recognizing}
Wilson, T.; Wiebe, J.; and Hoffmann, P.
\newblock 2005.
\newblock Recognizing contextual polarity in phrase-level sentiment analysis.
\newblock In {\em ACL Conf on Human Language Techn \& Empirical Methods in
  NLP},  347--354.

\bibitem[\protect\citeauthoryear{Yang \bgroup et al\mbox.\egroup
  }{2014}]{yang2014uncovering}
Yang, Z.; Wilson, C.; Wang, X.; Gao, T.; Zhao, B.~Y.; and Dai, Y.
\newblock 2014.
\newblock Uncovering social network sybils in the wild.
\newblock {\em ACM Trans. Knowledge Discovery from Data} 8(1):2.

\end{thebibliography}

\end{document}